\documentclass[twocolumn,prd,showpacs,preprintnumbers,amsmath,amssymb]{revtex4}
\usepackage{amsmath,mathrsfs,amsbsy,color,graphicx,bm,amsthm,amsfonts}
\usepackage{units}
\usepackage{bbm}
\usepackage{times}
\usepackage{dcolumn}
\usepackage{mathrsfs}
\usepackage{amsmath,amssymb,epsfig}

\begin{document}

\title{Influence of relativistic effects on satellite-based clock synchronization}
\author{Jieci Wang$^{1,2}$\footnote{ jcwang@hunnu.edu.cn}, Zehua Tian$^{1}$, Jiliang Jing$^{1}$ and Heng Fan$^{2}$\footnote{ hfan@iphy.ac.cn}}
\affiliation{
$^1$ Department of Physics, Collaborative Innovation Center for Quantum Effects, and Key Laboratory of Low
Dimensional Quantum Structures and Quantum\\
Control of Ministry of Education,
 Hunan Normal University, Changsha, Hunan 410081, People's
Republic
of China\\
$^2$ Beijing National Laboratory for Condensed Matter Physics, Institute of Physics,Chinese Academy of Sciences, Beijing 100190, People's
Republic
of  China}

\begin{abstract}
Clock synchronization between the ground and satellites is a fundamental issue in future quantum telecommunication, navigation, and global positioning systems. Here, we propose a scheme of near-Earth orbit satellite-based quantum clock synchronization with
atmospheric dispersion cancellation by taking into
account the spacetime background of the Earth. Two frequency entangled
pulses are employed to synchronize two clocks, one at a ground station and the
other at a satellite. The time discrepancy of the two clocks is
introduced into the pulses by moving mirrors
and is extracted by measuring the coincidence rate of the pulses in the interferometer.
We find that the pulses
are distorted due to effects of gravity when they propagate between
the Earth and the satellite, resulting in remarkably affected coincidence rates.
We also find that the precision of the clock synchronization is sensitive to
the source parameters and the altitude of the satellite.
The scheme provides a solution for satellite-based quantum clock synchronization
with high precision, which can be realized, in principle, with current technology.

\end{abstract}
\pacs{03.67.Hk, 04.20.-q, 06.30.Ft, 42.50.-p}

\vspace*{0.5cm}
\maketitle
\section{Introduction} High precision synchronization of clocks
plays an important role in modern society and scientific research \cite{synapp,synapp1}; examples include navigation, global
positioning, tests of general relativity theory, long baseline interferometry in radio
astronomy, as well as gravitational wave observation. Two standard classical protocols
for clock synchronization are  Einstein's  synchronization scheme \cite{einstsyn}
and Eddington's slow clock transfer \cite{Eddington}. The former requires
operational exchange of light pulses between the distant clocks and the latter is
based on sending a locally synchronized clock from one part to
other parts.  Recently, quantum strategies have been exploited to improve the
accuracy of clock synchronization.  A few quantum clock synchronization (QCS)
 proposals \cite{paper,qsyn1,qsyn2,qsyn3,qsyn4,qsyn41,qsyn5,qsyn6,yuefan1,yuefan2,qsyn71,qsyn72} and
 experiments \cite{qsyn8} are reported. It is shown that the schemes based
 on quantum mechanics
can gain significant improvements in precision
over their classical counterparts. 

On one hand several satellite-based quantum optics experiences \cite{Rideout12} are feasible with
current technology, such as satellite quantum communication \cite{Bonato, Vallone, Bonato1, Bruschi:Ralph:14,Erven, Rarity, Wangqkd,  Kurtsiefer, Villoresi}, and quantum tagging
\cite{Kent2},  as well as  gravity probes using beam interferometers \cite{Colella, Brodutch} and atomic clocks \cite{Kleppner,Alclock}  to test the principle of equivalence. Among these experiments, the  synchronization of clocks between a satellite and a ground station \cite{Giovannetti}
is an essential step.
Besides, a satellite-based quantum network of clocks is promising to act as a single world clock with
unprecedented stability and accuracy approaching the limit set by quantum mechanics, and there is also a security advantage \cite{Komar14}.
On the other hand, time dilation \cite{Alclock} is a concern because
of relativistic effects of the Earth on the QCS. The influence of relativistic effects on quantum systems \cite{RQI00, RQI0,RQI01,RQI02,RQI1,RQI2,RQI3,RQI4,RQI5} is a focus of study in recent years because such studies provide insights into some key questions in quantum mechanics and relativity, such as nonlocality, causality, and the information
paradox of black holes. Relativistic effects are particularly significant for the quantum versions of the
Eddington scheme \cite{paper,qsyn2,qsyn3}
because one must assume that the transfer is performed ``adiabatically slowly'' \cite{qsyn82} and the spacetime is
flat such that relativistic effects are negligible.
However, time dilation induced by Earth's spacetime curvature is experimentally observed for a change in height of $0.33$m \cite{Alclock} and thus cannot be neglected.

In this paper we propose a practical scheme for satellite-based QCS.
We let two observers, Alice and Bob, exchange
two frequency entangled pulses between a ground station and a satellite.
The influence of gravitational red-shift on the frequency of a pulse  can be eliminated by an opposite gravitational blue-shift. By assuming the clocks
have the same precision \cite{qsyn1,qsyn2,qsyn3},  clock synchronization can be realized by
identifying the time discrepancies.  In our scheme,
the time discrepancies are introduced through adding or subtracting optical path differences
into the pulses and the coincidence rate of the pulses in the interferometer as a function
of the time discrepancy between the clocks is measured. In actual
satellite-based quantum information processing
tasks \cite{Bonato,Wangqkd,Vallone,Villoresi, Erven, Bruschi:Ralph:14, Bruschi:Ralph:15},
and similarly in protocols of QCS, the main
errors are induced by photon-loss and the
dispersion effects of the atmosphere through which
the pulses travel. Therefore, we employ  frequency entangled light, instead
of the entangled N00N state which is vulnerable for photon-loss \cite{Giovannetti11},
as well as the dispersion cancellation
technology \cite{kwiat1,kwiat2,qsyn3} to eliminate the influence of the atmospheric
scattering.  We find that the coincidence rate of interferometry is remarkably affected by the spacetime
curvature of the Earth. We also find that the precision of the clock synchronization is sensitive to
the light source parameters.

 The outline of the paper is as follows. In Sec. II we briefly  introduce the
sketch of the experimental setup. In Sec. III we discuss how the Earth's spacetime effects
the propagation of  photons. In Sec. IV we study the feasibility of satellite-based QCS
and how the effect of the Earth's  gravity will disturb it. In the last section
we discuss the experimental feasibility of our scheme and give a  brief summary.

\begin{figure*}[tbp]
\centerline{\includegraphics[width= 15 cm]{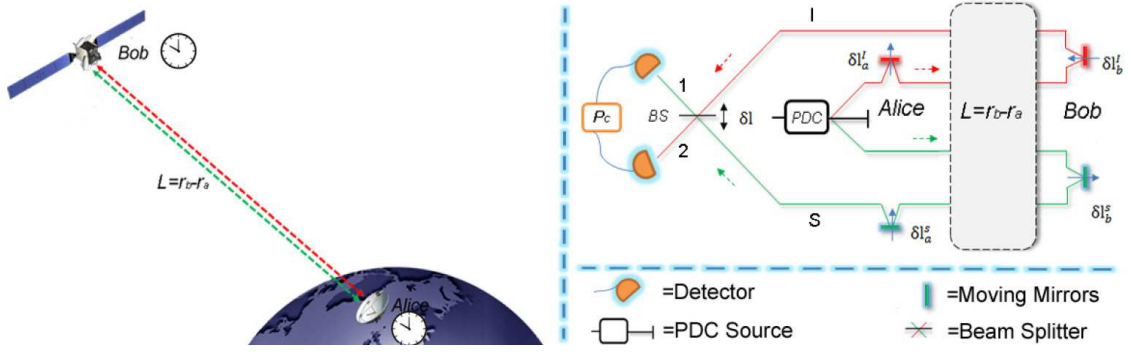}}
\caption{Sketch of the experimental setup for satellite-based QCS. Alice sends two frequency entangled beams produced by a parametric down converter crystal (PDC) source to Bob who bounces them back to Alice again. Alice \emph{adds} an OPD $\delta l^I_a(t)$ to the idler beam \cite{qsyn3} and \emph{subtracts } the same amount of OPD $\delta l^S_a(t)$ from the signal beam which was bounced back from Bob. At the satellite Bob \emph{subtracts} an OPD $\delta l^I_b(t)$ from the idler
beam and \emph{adds} an identical OPD $\delta l^S_b(t)$ to the signal beam.  Finally, we let the pulses interfere at the 50/50 beam splitter (BS) and be detected at the measurement setups.  The coincidence rate $P_c$ is measured as a function
of the time discrepancy $\tau_0^b-\tau_0^a$ between the clocks.}
\label{f:experiment}
\end{figure*}

\section{The scheme} The sketch of our proposal for satellite-based QCS is described in Fig. (\ref{f:experiment}).
The quantum optical technology
of our proposal is based on the Hong, Ou, and Mandel (HOM)
interferometer \cite{manou}. We assume that Alice works on the surface
of the Earth ($r=r_A$) with her own clock, while Bob works on a satellite
at constant radius $r=r_B>r_A$.  The clocks have the same accuracy thus the
clock synchronization problem is reduced to the problem of identifying the time
discrepancy between the clocks. Alice sends two frequency entangled beams produced by a parametric down converter crystal (PDC) source to Bob and Bob bounces them back to Alice again.
Those two pulses are named signal $(S)$ beam and idler $(I)$
beam, respectively. By exchanging entangled beams between Alice and Bob, a ``conveyor belt" \cite{qsyn3} for time information is established.  After propagating
through different optical paths, the signal and idler
beams are interfered at the 50/50  beam splitter and be measured by the detectors. To introduce time information into the beams,  Alice and Bob use  moving mirrors
with constant speed $v$ to add or subtract  optical path differences (OPD) to the beams \cite{qsyn3}. Alice and Bob come to an agreement on the  starting  time $\tau_0$ of their mirrors in advance. Since they do not have a synchronized
clock to start with, they  can only start the mirrors at time $\tau_0^i (i=a,b)$  relative to the time readings of their own
clock, which are different due to different locations.  As described in  Fig. (1),  Alice at the ground station start a moving mirror
to \emph{add} an OPD $\delta l^I_a(t)$ to the idler beam \cite{qsyn3} and to \emph{ subtract }  the same amount of OPD $\delta l^S_a(t)$ from the signal beam which was bounced back from Bob. At the satellite Bob \emph{subtracts} an OPD $\delta l^I_b(t)$ from the idler
beam and \emph{adds} an identical OPD $\delta l^S_b(t)$ to the signal beam.   The linear time
dependent OPDs are given by
\begin{eqnarray}
\begin{array}{ll}
\delta l^I_a(t)=v(t-\tau_0^a),\qquad&
\delta l^I_b(t)=-v(t-\tau_0^b),\\\\
\delta l^S_a(t)=-v(t-\tau_0^a),\qquad &
\delta l^S_b(t)=v(t-\tau_0^b)\;,\end{array}
\;\label{delayt}
\end{eqnarray}
where $\tau_0^a$ and $\tau_0^b$ are the starting times (proper time) as measured by
Alice's and Bob's clocks, respectively. In Eq. (\ref{delayt}) $t$ is the time point (coordinate time) of the coincidence detection of the signal and idler
photons, i.e., the time when the photons quantum state is measured. By assuming the quantum state is instantaneously collapsed by the measurements made on the surface of the Earth, we can agree that the collapsing time is identical $t=t^a=t^b$, even if Alice and Bob' reading times are different $\tau^a\neq \tau^b$ at this moment.  From Eq. (\ref{delayt}) we can see that the delays are proportional to the time interval between the starting time $\tau_0^i$ for the moving of the mirrors and the time $t$  when the photons are detected. If the proportionality
constant  $v$ in Eq. (\ref{delayt}) and the starting time reading $\tau_0^i$ on Alice's and Bob's clocks are  identical,  the quantity
of OPD at Alice's point  will be \emph{zero} after an exchange period.
Therefore, the final OPD is totally produced from the starting time discrepancy $\Delta\tau= \tau_0^b-\tau_0^a$ between  Alice's clock and Bob's
clock. Then the signal and idler
pulses are interfered at the beam splitter (BS) and click the detectors.
We will show in section IV that the final difference of optical path lengths is affected by a factor depending on the starting
time discrepancy $\Delta\tau$.   By measuring
the photon coincidence rate $P_c$ at the output ports 1 and
2 of the beam splitter, one may acquire very precise
information on the OPD in the two
arms.  Thus, it is
sufficient to measure the photon coincidence rate to recover
the exact time discrepancy between Alice's clock and Bob's
clock. Then Alice tells Bob
by classical communication to adjust his clock  according to the time discrepancy.
By using this scheme, Alice and Bob may check  how much the accuracy is
disturbed by the gravity induced spacetime curvature of the Earth, and inversely they can precisely measure the curvature via an atomic clock, which is the most accurate setup currently available in the world.

\section{Earth's spacetime on the propagation of photons}
Now we describe the propagation of photons from the Earth to a satellite
by taking gravity of the Earth into consideration.
The Earth's spacetime curvature will influence the light pulses during
their propagation between the ground
station and the satellite.  We know that the Earth rotates slowly with an angular velocity at the
equator of $\omega_E = 7.2921151247\times 10^{-5} rad/ s$ or at a linear speed of $v_E =465\ m s^{-1}$,
 which is much slower than the speed of light. Therefore,  the Schwarzschild metric \cite{Liao14,Misner:Thorne:73,Wald:84} is a sufficient
 approximation for the Earth's spacetime, as has been discussed in \cite{Bruschi:Ralph:14, Bruschi:Ralph:15,Liao14}.  The Schwarzschild metric is given by \cite{Misner:Thorne:73,Wald:84}
\begin{eqnarray}
ds^2&=&-(1-\frac{r_s}{r})dt^2+(1-\frac{r_s}{r})^{-1}dr^2\nonumber\\&&+r^2(d\theta^2
+sin^2\theta d\varphi^2),
\end{eqnarray}
where  $r_S:=\frac{2GM}{c^2}$ is the Earth's Schwarzschild radius, $M$ is the mass of the Earth, $c$ is the
speed of light in vacuo, and $G$ is the gravitational constant.

A photon can be properly modeled by a wave packet of  electromagnetic fields \cite{Bruschi:Lee:13}
with a distribution $F^{(K)}_{\omega_{K,0}}$ of modes peaked around the
frequencies  $\omega_{K,0}$ \cite{Downes:Ralph:13,Leonhardt:05}, where $K=A,B$ labels
either Alice or Bob. The annihilation operator for a photon from the point of view of
Alice or Bob takes the form
\begin{eqnarray}
a_{\omega_{K,0}}(t_K)=\int_0^{+\infty}d\omega\Omega_K\, e^{-i\omega_K t_K} F^{(K)}
_{\omega_{K,0}}(\omega_K)\,a_{\omega_K},\label{wave:packet}
\end{eqnarray}
where  $\omega_K$ is the physical frequencies as measured in their labs. 
At time $\tau_A$ Alice prepares a wave packet $F^{(A)}_{\omega_{A,0}}$ and sends it to Bob who will receive it at a different proper time $\tau_B=\Delta\tau+\sqrt{f(r_B)/f(r_A)}\tau_A$, where $f(r_A)=1-\frac{r_s}{r_A}$, $\Delta\tau$ is propagation time of the wave packet, and the factor $\sqrt{f(r_B)/f(r_A)}\tau_A$ indicates the relativistic time dilation.
The wave packet received is modified due to the spacetime curvature of the
Earth and takes the form $F^{(B)}_{\omega_{B,0}}$.
By employing the definition of proper time, it is easy to show that the time evolution
for the modes has the form $i\partial_{\tau_K}\phi_{\omega_K}=\omega_K\phi_{\omega_K}$,
where $\phi^{(u)}_{\omega_K}$ are the quantum states of the modes corresponding to the
operators $a_{\omega_{K}}$ \cite{Bruschi:Ralph:14}. This equation indicates that the
physical frequency $\omega_K$ measured by an observer at the position $r_K$ is $\omega_K=f(r_K)^{-1/2}\omega$.
Then we find that Bob will receive a mode with frequency
$\omega_B=\sqrt{\frac{f(r_A)}{f(r_B)}}\omega_A$
 if a sharp frequency mode with $\omega_A$ was sent by Alice.
This is the well-known gravitational red-shift effect, which was predicted by Einstein in  1911 and experimentally verified in 1960 \cite{Pound1960}.  In our scheme such a classical gravitational effect is designed to be eliminated by an opposite gravitational blueshift
factor $\sqrt{\frac{f(r_B)}{f(r_A)}}$ because the signal will be sent downward from a satellite to the Earth.

However, such a nonlinear gravitational effect is found to influence the fidelity of the quantum channel between Alice and Bob \cite{Bruschi:Ralph:14,Bruschi:Ralph:15} and will inevitably affect the accuracy of the satellite-based QCS as well. The mode $\bar{a}^{\prime}$ received by Bob can be
decomposed in terms of the mode $a$ prepared by Alice
and an orthogonal mode $a_{\bot}$ \cite{Bruschi:Ralph:14,Bruschi:Ralph:15,Rohde:Mauerer:07}
\begin{eqnarray}
\bar{a}^{\prime}=\Theta\,a+\sqrt{1-\Theta^2}\,a_{\bot},\label{mode:decomposition}
\end{eqnarray}
where  $\Theta$ is the wave packet overlap between the
distributions $F^{(B)}_{\omega_{B,0}}(\omega_B)$ and $F^{(A)}_{\omega_{A,0}}(\omega_B)$,
\begin{eqnarray}
\Theta=\int_0^{+\infty}d\omega_A\,F^{(A)\star}_{\omega_{A,0}}
(\omega_B)F^{(B)}_{\omega_{B,0}}(\omega_B),\label{single:photon:fidelity}
\end{eqnarray}
which describes the fidelity of the channel between Alice and Bob. For a perfect channel one has $\Theta=1$.

\section{ Spacetime curvature on satellite-based QCS}
 The emitted signal and idler beams from the PDC initially share an entangled
state \cite{qsyn3}
\begin{equation}
|\psi\rangle = \int d\omega_1 d\omega_2
F(\omega_1,\omega_2) a^\dagger(\omega_1)
a^\dagger(\omega_2)|0\rangle,
\end{equation}
where $a^\dagger(\omega_1)$ and $a^\dagger(\omega_2)$  are creation operators  of the first and the second photons, respectively.
For a monochromatic pump this state can be rewritten as
\begin{eqnarray}
 |\psi\rangle=\int d\omega
\;F_{\omega_0}(\omega)\;|\omega_0+\omega\rangle_I|\omega_0-\omega\rangle_S
\;,
\end{eqnarray}
where $|\omega\rangle_S$ and $|\omega\rangle_I$ are the states for the signal and
idler pulses, and $F_{\omega_0}(\omega)$ is the spectral distribution function of the
down-converted photons \cite{kwiat1,kwiat2}. Now let us briefly discuss how quantum  entanglement is useful to quantum clock synchronization and the advantage of the quantum clock synchronization scheme. The main advantage of a quantum strategy is that we can employ quantum uncertainty and coherence time of the frequency entangled photons. From Eq. (7) we can see that
although the sum frequency $2\omega_0$ is
certain, the down-shifted
frequencies $|\omega_0-\omega\rangle_S$ and $|\omega_0+\omega\rangle_I$ are highly uncertain.  The frequencies
are largely determined by the pass bands of the interference
filters  that inserted in the down-shifted beams \cite{manou}. These pass bands have been found on the order of
$5\times10^{12}$ Hz, which corresponds to a coherence time for each
photon on the order of 100 fs \cite{manou}. Therefore, it is able to measure a time interval of or better than the coherence time of the photons (50 fs), with an accuracy of 1 fs ($10^{-15}s$). Also, the most important task for the quantum clock synchronization is to determine the time interval by measuring the
rate at which photons are detected in coincidence, which relates to the coherence length of
the photon wave packet and entanglement of the photons.

 As stated before,  Alice \emph{adds} an OPD $\delta l^I_a(t)$ to the idler beam \cite{qsyn3} and \emph{subtracts } the same amount of OPD $\delta l^S_a(t)$ from the signal beam which was bounced back from Bob. At the satellite Bob\emph{subtracts} an OPD $\delta l^I_b(t)$ from the idler
beam and \emph{adds} an identical OPD $\delta l^S_b(t)$ to the signal beam. Then we let the pulses
interfere at the 50/50  BS and be measured in the HOM interferometer. The detected coincidence rate $P_c$ at the detectors is given by the Mandel formula {\cite{mandel}}
\begin{eqnarray} &&P_c\propto
\int_t dt_1dt_2\;\langle\psi|\mathcal{E}_1^{(-)}\mathcal{E}_2^{(-)}
\mathcal{E}_2^{(+)}\mathcal{E}_1^{(+)}|\psi\rangle
\;,\label{mand}
\end{eqnarray}
where $t$ is the interaction time of the detectors, and the electromagnetic fields at time $t_j$ at the
output of the beam splitter can be defined by
\begin{eqnarray}
\nonumber\mathcal{E}_j^{(+)}&=&i\int
d\omega\;\sqrt{\frac{\hbar\omega}{4\pi \zeta c}}\;a_j(\omega)
e^{-i\omega(t_j-x_j/c)}\\&=&\left(\mathcal{E}_j^{(-)}\right)^\dag, \mbox{for }j=1,2\;,
\label{campoe}
\end{eqnarray}
where $\zeta$ is the beam cross section,  and $x_j$ is the position of the  moving mirrors
detector. The relationship between the output and input fields of the beam of the moving mirrors
can be obtained by performing Lorentz transformations on
the input fields \cite{qsyn3}. For example, the factor $a_j(\omega)e^{-i\omega(\tau_0^j-x_0^j/c)}$ after
the beams outflow from the moving mirrors evolves
to $\sqrt{\chi}a_j(\chi\omega)e^{-i\omega\frac{2\beta}{1-\beta}(\tau_0^j-x^j_0/c)}$,
where $\beta=\frac vc$ is the Lorentz transformation factor,  $\chi = \frac{1+\beta}{1-\beta}$
denotes the Doppler shift introduced by the moving mirrors, and $x_0^j$ is
the distance of the beam's delay. For the time delays defined in Eq. (1), we have $x_0^j=x_0$ for all $j$. Then the beams are sent from the Earth to the satellite and are influenced
by the dissipation of the atmosphere and the spacetime curvature. The former produces a phase
discrepancy $i\kappa^j_t(\omega)$ and the latter can be described by Eq. (5).
Taking the signal beam as an example, the full procedure can be expressed by the
transformations between the annihilation operators when the photons
are sent from the Earth to the satellite
\begin{eqnarray}\label{doppsign}
 a'_S(\omega)&=&\Theta_{1}\sqrt{\chi}\; a_S\bigg(\chi\sqrt{\frac{f(r_A)}{f(r_B)}}\omega\bigg)
\;e^{{-i\sqrt{\frac{f(r_A)}{f(r_B)}}\omega\Upsilon_1+i\kappa^S_t(\omega)
}}\nonumber\\ &&+\sqrt{1-\Theta_1^2}a^{\perp}_S\bigg(\chi\sqrt{\frac{f(r_A)}{f(r_B)}}\omega\bigg),
\end{eqnarray}
where $\Upsilon_1=\frac{2\beta}{1-\beta}(\tau_0^b-x_0/c)- L/c$ and $\Theta_{1}$ is the wave packet overlap
between the distributions when the pulse is sent from the
Earth to the satellite.
We can see that the annihilation operator $a_S(\omega)$ at Alice's laboratory
evolves into $a'_S(\omega)$ when observing at Bob's laboratory, where  $L$ is the distance between the
Earth and the satellite. Again, the annihilation operator for the signal beam evolves into $a''_S(\omega)$
before entering the 50/50 BS
\begin{eqnarray}\label{doppsign1}
a''_S(\omega)&=&\Theta_{1}\Theta_{2}a_I(\omega)
e^{{i\omega\Upsilon_2+
i\kappa^S_t(\omega)+i\kappa^S_f(\omega/\chi)}}\nonumber\\&&+\Lambda a^{\perp}_S(\omega)
\;,
\end{eqnarray}
where $\Upsilon_2=\frac{2\beta}{1+\beta}(\tau_0^a-\tau_0^b+
\frac{x_0}c)+(L/\chi+ L')/c$, $\Lambda=\Theta_{2}\sqrt{1-\Theta_1^2}+\sqrt{1-\Theta_2^2}$, $L'$ denotes the distance between Bob and the BS, and  $\Theta_{2}$ is
the mode overlap between the distributions when the pulse is received from the satellite.
In Eq. (\ref{doppsign1}) the terms $\kappa^S_t$ and $\kappa^S_f$ describe  the effect of the
dispersive atmosphere on the signal beam on
their way  to and  from the satellite, respectively. Notice that
the Doppler shift  introduced by the first mirror is $\omega/\chi$ and the frequency itself is
revaluated at $\omega$ again because the second mirror moves in the opposite direction \cite{qsyn3}.
The analogous procedures can be applied to the
idler beam propagation process, yielding  the final operator transformations
\begin{eqnarray}\label{dopplidler}
\nonumber a''_I(\omega)=\Theta_{1}\Theta_{2}a_I(\omega)
e^{{-i\omega\Upsilon_3+
i\kappa^I_t(\omega/\chi)+i\kappa^I_f(\omega)}}+\Lambda a^{\perp}_I(\omega)
\;,
\end{eqnarray}
where $\Upsilon_3=\frac{2\beta}{1+\beta}(\tau_0^a-\tau_0^b-
\frac{x_0}c)-(L/\chi+ L')/c$.
Because the distance between the BS and the PDC source is much smaller than the distance between the satellite and the Earth, we assume that $L=L'$.
At the output of the 50/50 BS, the modes are found to be
$a_1(\omega_1)=\frac 1{\sqrt{2}}
[ia''_I(\omega_1)\;e^{-i\omega_1\delta l/c}
+a''_S(\omega_1)]
$ and $a_2(\omega_2)=\frac 1{\sqrt{2}}
[ia''_S(\omega_2)+a''_I(\omega_2)\;
e^{-i\omega_2\delta l/c}]$,
where $\delta l$ is the delay introduced to relate the coincidence rate $P_c$
with the path length. Then the coincidence rate $P_c$ defined in  (\ref{mand}) is obtained as
\begin{eqnarray} P_c&&\propto\int d\omega_1d\omega_2\langle
\psi|a^{\dagger}_1(\omega_1)a^{\dagger}_2(\omega_2)a_1(\omega_1)a_2(\omega_2)|\psi\rangle \nonumber\\ &&=\int d\omega_1d\omega_2|\langle
0|a_1(\omega_1)a_2(\omega_2)|\psi\rangle|^2
\;\label{coinrate}.
\end{eqnarray}
The  matrix element $\langle 0|a_1(\omega_1)a_2(\omega_2)|\psi\rangle$ in (\ref{coinrate}) is given by
\begin{eqnarray} &&\langle 0|a_1(\omega_1)a_2(\omega_2)|\psi\rangle=
\frac 12\;(\Theta_{1}\Theta_{2})^2\delta(\omega_1+\omega_2-2\omega_0)
\;
\nonumber\\&&
e^{i\varphi}\phi({\omega_1}-\omega_0)
\Big[1-e^{{2i({\omega_1-\omega_0})\Upsilon_4
-i\Delta\kappa(\omega_1)}}\Big]
\label{panino}\;,
\end{eqnarray}
where $\Upsilon_4=\frac{4\beta}{1+\beta}(\tau_0^b-\tau_0^a)-\delta
l/c$, and $\varphi$ is an overall phase term that will disappear by taking
the modulus \cite{qsyn3}, and the contribution of the dispersion terms is
\begin{eqnarray}
\nonumber&&\Delta\kappa(\omega)=\kappa^S_t(\omega)-\kappa^I_f(\omega)+
\kappa_f^I(\omega') -\kappa_t^S(\omega')
\\&&+\kappa_t^I(\omega')
-\kappa_f^S(\frac{\omega_1}{\chi})+
\kappa^S_f(\frac\omega\chi)-\kappa^I_t(\frac\omega\chi),
\;\label{deltak}
\end{eqnarray}\
where $\omega'=2\omega_0-\omega$.
If the properties of the
beams propagating
through different optical paths are such that $\kappa_t^S=\kappa_f^I$ and $\kappa_f^S=\kappa_t^I$,  the dispersion  effect of the atmosphere is erased.
Such  conditions can be satisfied by
allowing the ``from" idler beam to propagate at a distance
less than the spatial inhomogeneities of the atmosphere from
the ``to" signal beam and, equivalently, by allowing the ``to"
idler beam to propagate less than the ``from" signal beam \cite{qsyn3}. Here, the ``from" beam denotes the beam from Alice to Bob and vice versa. Then the dispersion suffered
by one of the photons can cancel that suffered
by the other photon. These two photons can remain
totally coincident after propagating
through different optical paths. Substituting Eq. (\ref{panino}) into Eq. (\ref{coinrate}), we can obtain
\begin{eqnarray}
P_c=(\Theta_{1}\Theta_{2})^2\int d\omega |\mathcal{F}|^2(1-\cos[\frac {2\omega}
c(\frac{4v\Delta\tau}{1+\beta}-\delta l)]),
\;\label{fin}
\end{eqnarray}
where $\mathcal{F}=F_{\omega_0}(\omega)$.
From Eq. (\ref{fin}) we can see that the coincidence rate $P_c$ is directly related
to the time discrepancy $\Delta\tau=\tau_0^b-\tau_0^a$ of Alice's and Bob's clocks.
That is to say, Alice's and Bob's clocks can be synchronized
by using the measured coincidence rate of interferometry.

To be explicit we next only consider the case in which $F_{\omega_0}(\omega)$ is a Gaussian wave packet
$F_{\omega_0}(\omega)=\frac{1}{\sqrt[4]{2\pi\sigma^2}}e^{-\frac{(\omega-\omega_0)^2}{4\sigma^2}}$,
where $\sigma$ is the Gaussian width. The wave packet overlaps $\Theta_1$ and $\Theta_2$ are found to be
\begin{eqnarray}
\label{fintt}\Theta_{1(2)}=\sqrt{\frac{2\Delta_{1(2)}}{1+\Delta_{1(2)}^2}}
e^{-\frac{\vartheta^2\omega_{0}^2}{4\sigma^2[1+\Delta_{1(2)}^2]}}\label{final:result},
\end{eqnarray}
where $\Delta_{1(2)}=1\pm\vartheta$ and the signs $\pm$ occur for $r_A<r_B$ or $r_A>r_B$.  In Eq. (\ref{fintt}), we define $\vartheta=\sqrt{\frac{f(r_A)}{f(r_B)}}-1$ and
$\omega_{A,0}=\omega_{B,0}=\omega_0$ is assumed.
The modes will be perfectly overlapped ($\Theta=1$)
when Alice and Bob are in a flat spacetime $f(r_A)=f(r_B)=1$. 
For the typical resources used in quantum optics experiments, the relation  $\vartheta\ll(\frac{\vartheta\omega_{0}}{\sigma})^2\ll1$ should be
satisfied \cite{Bruschi:Ralph:14, Bruschi:Ralph:15,Matsukevich:Maunz:08}, which yields
 $\Theta_1=\Theta_2\sim1-\frac{\vartheta^2\omega_{0}^2}{8\sigma^2}$.  Then we find that
the coincidence rate $P_c$ has the form
\begin{eqnarray} &&P_c= (1-\frac{\vartheta^2\omega_{0}^2}{8\sigma^2})^4[1-e^{-2\sigma^2\;
(\delta l-\frac{4v\Delta\tau}{1+\beta})^2/c^2}]
\;\label{gauss}.
\end{eqnarray}
We can see that the coincidence rate $P_c$ has the factor
$(1-\frac{\vartheta^2\omega_{0}^2}{8\sigma^2})^4$ compared to that of the flat spacetime case,
where $P^{f}_c= 1-e^{-2\sigma^2\;(\delta l-\frac{4v\Delta\tau}{1+\beta})^2/c^2}$ \cite{qsyn3}. We define the effect of spacetime curvature on the accuracy of clock synchronization as the relative disturbance
of coincidence rate
\begin{eqnarray}
\Delta_p=\frac{P^{f}_c-P_c}{P^{f}_c}=1-(1-\frac{\vartheta^2\omega_{0}^2}{8\sigma^2})^4.
\end{eqnarray}
It is now clear that the relative disturbance of coincidence rate $\Delta_p$ depends on
the spacetime parameter $\vartheta$ and the characteristics of the PDC source.

\begin{figure}
\includegraphics[scale=0.9]{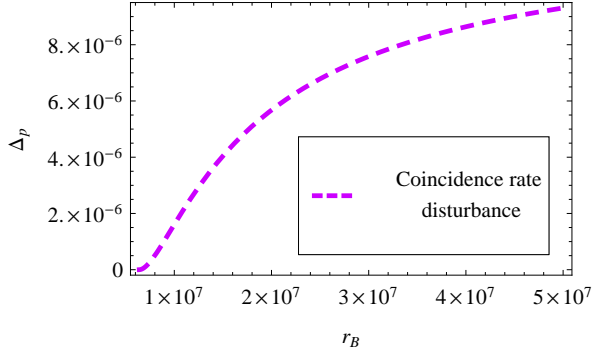}
\caption{(Color online) The relative disturbance of coincidence rate $\Delta_p$ as a function of the altitude of the satellite $r_B$. The parameters of the PDC  light source are fixed as $\omega_0=700$THz and $\sigma=100$MHz.}\label{Fig2}
\end{figure}

In Fig. (\ref{Fig2}), we plot the relative disturbance of the Earth's spacetime curvature on the coincidence rate $\Delta_p$ as a function of the distance $r_B$ between the satellite and the Earth's core for the fixed light source parameters. It is shown that $\Delta_p$ increases as the distance increases, i.e.,  the accuracy of clock synchronization depends on the altitude of the satellite, which also verifies that the spacetime curvature remarkably influences the  running of the atomic clocks \cite{Alclock}. This result is very different from that of Ref. \cite{qsyn3}, in which the coincidence rate is
independent of the distance $L$ between Alice and Bob when the spacetime curvature of the Earth is not considered.

\begin{figure}
\includegraphics[scale=0.56]{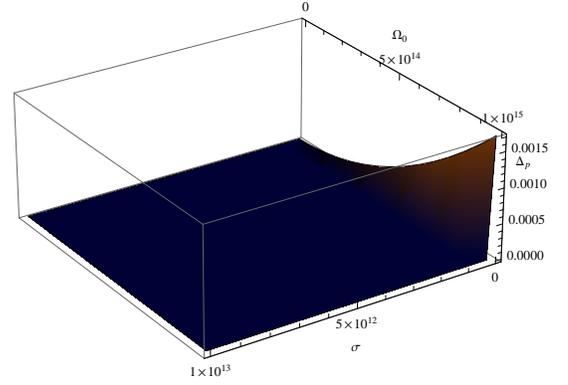}
\caption{(Color online) The relative disturbance of coincidence rate $\Delta_p$ as a function of the source parameters $\omega_0$ and $\sigma$  for the fixed distance $r_B=6.771\times10^6$m (LEO).}\label{Fig3}
\end{figure}

In Fig. (\ref{Fig3}), we plot the relative disturbance  $\Delta_p$ over the peak frequency $\omega_0$ and bandwidth $\sigma$ of the PDC. It is shown that the disturbance of accuracy depends sensitively on the bandwidth $\sigma$ of the source, which is similar to the flat spacetime case \cite{qsyn3}. However, here we find that the disturbance on accuracy also depends on the  peak frequency  of the pulses, which is different from that of \cite{qsyn3}
where the accuracy is independent of the peak frequency. In this paper we are particularly interested in two typical cases, in which the QCS are performed between
the ground station and either a low earth orbit satellite (LEO), or a geostationary earth
orbit (GEO) satellite, respectively.

\emph{The LEO case:}
The typical distance from the Earth to a LEO satellite is about $400$km,
which yields $r_A=6.371\times 10^6$m and $r_B=6.771\times 10^6$m. Considering that the
Schwarzschild radius of the Earth is $r_S=9$mm, it is found that
$\delta\sim-\frac{1}{2}(\frac{r_s}{r_B}-\frac{r_s}{r_A})=4.17\times 10^{-11}$. We employ a typical PDC source with a
wavelength of 369.5 nm (corresponding to $\omega_{0}=812$THz) and $\sigma=100$MHz (the relation $\delta\ll(\frac{\delta\omega_{0}}{\sigma})^2\ll1$ is
satisfied).
A light source with such peak frequency and bandwidth
is available, for example, in trapped ion experiments \cite{Matsukevich:Maunz:08}.  The
relative disturbance of the spacetime curvature on the coincidence rate
is obtained as $\Delta_p^L=5.73993\times10^{-8}$.  The achievement of an optical lattice clock
with accuracy at the $10^{-18}$ level has been reported in Refs. \cite{Hinkley,Bloom}.
If we would like to synchronize two clocks up to a time discrepancy of $\tau=1$ns (as in \cite{qsyn3},
or a much lower level $\tau=100$ns as presented in \cite{qsyn2}), the correction of the
Earth's spacetime curvature effect will reach $10^{-17}$s during a single synchronization process.
Such a correction is comparable to
the  accuracy of the atom clocks and thus should be considered for the QCS between clocks
in  future satellite-based applications.  Therefore, we can safely arrive at the conclusion
that  the spacetime curvature is {\it not negligible} when the synchronization is performed by LEO satellites.

\emph{The GEO case:}
The typical distance $L$ between a GEO satellite and the ground is about $3.6\times10^7$m. Therefore,
the distance between the Earth and the satellite $r_B$ is about $r_B = 42.371\times10^6$m, which yields $
\delta\sim-\frac{1}{2}(\frac{r_s}{r_B}-\frac{r_s}{r_A})=6\times 10^{-10}$.
In this case the relative disturbance of the spacetime curvature on the coincidence rate is $\Delta_p^G=1.18729\times 10^{-5}$. We
find that the disturbance of the spacetime curvature on the coincidence rate for the GEO satellites becomes even {\it more remarkable} than that of the LEO
case.
We remark that the current GPS satellites have $r_B \approx 2.7\times10^7$m. In the QCS scheme, the
spacetime curvature is also remarkable.

\emph{Error analysis:} It is to mentioning that the velocity variations of the moving mirrors may induce some  errors  on the coincidence rate. However, the order of magnitudes of the movement speed of the mirrors is much smaller than the velocity of light, let alone the  velocity variation of the mirrors. To be specific, the typical velocity of the moving mirrors is $10^{-1}$ \emph{m/s}. Let us suppose that this velocity has one percent of variation, say $10^{-3}$ \emph{m/s}, which is much smaller than the velocity of light. Note that the relative disturbance of the spacetime curvature on the coincidence rate is on the order of $10^{-8}$ for the LEO satellites and of the order of $10^{-5}$ for the GEO satellites, which are at least 3 orders of magnitudes larger than that of the velocity variations of the mirrors.  Therefore, the systematic errors induced by the velocity variation of the mirrors can be safely ignored in the scheme.

\section{ Discussions} We have proposed a practical
satellite-based QCS scheme with the advantages of dispersion cancellation
and the robust frequency entangled pulses of light by
taking the effects of the spacetime curvature of the Earth into consideration. The spacetime
background of the Earth is described by the Schwarzschild metric, and the quantum optics part of our
proposal is based on the HOM interferometer.
By eliminating the gravitational redshift and blueshift of the laser pulses
and the atmospheric dispersion cancellation, the accuracy of  clock
synchronization in our quantum scheme can be very high by showing that $\Delta _p$
is close to unity.
Our proposal can be implemented, in principle, with current available
technologies.  To be specific, optical sources with the required peak frequency and bandwidth have been
achieved by the trapped ion experiments \cite{Matsukevich:Maunz:08}. The feasibility of photon exchanges
between a satellite and a ground station has been experimentally demonstrated \cite{Villoresi} by the
Matera Laser Ranging Observatory (MLRO) in Italy. Most recently, they have reported the operation of
experimental satellite quantum communications \cite{Vallone}
by sending selected satellites laser
pulses. Our scheme can also be generalized to the quantum clock network cases \cite{Komar14,paper}. The results should be significant both for determining the accuracy
 of clock synchronization and for our general understanding of time discrepancy in future
 satellite-based quantum systems.

\begin{acknowledgments}
This work is supported by the National Natural Science Foundation
of China under Grants No. 11305058, No. 11175248, No. 11475061, the Doctoral Scientific Fund Project of the Ministry of Education of China under Grants No. 20134306120003, Postdoctoral Science Foundation of China under Grants No. 2014M560129, No. 2015T80146 and the Strategic Priority Research Program of the Chinese Academy of Sciences (under Grant No. XDB01010000).
\end{acknowledgments}

\end{document}